\documentstyle[11pt,aaspp4]{article}
\slugcomment{ The Astrophysical Journal }
\lefthead{Che et al.}
\righthead{Gamma-ray Bursts}

\begin{document}

\title{On Source Density Evolution of Gamma-ray Bursts}

\author{H. Che\altaffilmark{1}, Y. Yang\altaffilmark{2} and R. J. Nemiroff\altaffilmark{3}}
\affil{Department of Physics, Michigan 
Technological University,
Houghton, MI  49931}
\altaffiltext{1}{hche@mtu.edu}
\altaffiltext{2}{yuyang@mtu.edu}
\altaffiltext{3}{nemiroff@mtu.edu}
\begin{abstract}
Recent optical afterglow observations of gamma-ray bursts indicate a setting and distance scale that many relate to star-formation regions. In this paper, we use $<V/V_{max}>$ and a set of artificial trigger thresholds to probe several potential GRB source density evolutionary scenarios.  In particular, we compare a uniform subset of BATSE 4B data to cosmological scenarios where GRBs evolve as the comoving density, the star formation rate, the QSO rate, and the SN Type Ic rate.  Standard candle bursts with power-law spectra and a universe without vacuum energy were assumed.  Our results significantly favor a comoving density model, implying that GRB source density evolution is weaker than expected in these evolutionary scenarios.  GRB density might still follow star-formation rates given proper concurrent GRB luminosity evolution, significant beaming, significant error in standard candle assumptions, or were a significant modification of star formation rate estimates to occur.
\end{abstract}


\section{ Introduction }
Recent afterglow and counterpart observations of gamma-ray bursts (GRBs), particularly the redshift-related observations of GRB 970508, GRB 971214 and GRB 980703, have established the cosmological origin of gamma-ray bursts (GRBs) (Bloom et al. 1998, Kulkarni et al. 1998, Djorgovski et al. 1998, Galama et al. 1998).   What sources could create such high-energy outbursts with bulk energy of $> 10^{51}$?  The identification of [OII] emission lines in GRB 970508 and GRB 980703 strongly suggests that GRBs are associated with vigorous star formation galaxies (Metzger et al. 1997, Djorgovski et al. 1998).  The temporal-spatial coincidence of GRB 980425 and SN1998bw, a type Ic supernova, has created speculation that some GRBs are related to certain types of supernovae (Iwamoto et al. 1998, Woosley et al. 1998), although in the specific
case of GRB 980425, the gamma-ray flux would then be low.

Based on these observations, a variety of GRB models have been proposed, including 'hypernova', the collapse of rapidly rotating massive stars with strong magnetic fields ($B \sim 10^{15}$G) in a binary systems (Paczynski, 1997), and the merger of a helium star and black hole (Fryer \& Woosley 1998).   Models that tie GRBs to massive stars implicitly suggest that GRBs are related to star formation in some way.

On the other hand, the acceptable fit of the GRB luminosity distribution to the comoving constant rate in Friedman universe has been thought as a consistent result  of cosmological origin of GRBs prior to the discoveries of after-glows (Mao \& Paczynski 1992, Fenimore 1993, Piran 1992, Wickramasinghe et al. 1993). Previously, Totani (1997) and Marani (1998) have noted that a constant comoving rate for GRBs fits measured $\log N -\log P$ distributions significantly better than published rate estimates involving star formation.  Examining possible reasons for differences between GRB rates and star formation rates (SFRs) estimates is therefore important.  

In this paper, we make use of $<V/V_{max}>$ to trace source density evolution. 
Although $<V/V_{max}>$ was originally created to test homogeneity in the face of multiple detectors and detection criteria (Schmidt et al.1988), this statistic remains a sensitive function of homogeneity, and hence source density evolution, even when only a single detector is involved.  More specifically, we construct a set of sub-samples of BATSE GRBs by artificially changing the threshold of BATSE in the 4B Catalog.

Unlike the more traditional method of quantifying evolution by comparing observational and theoretical distributions of log $N - $ log $P$ with a $\chi^2$ or Kolmogorov-Smirnov test, the $<V/V_{max}>$ comparison we employ has no free parameters.  The spectral index is fixed by observations, and the evolutionary function is well defined by popular cosmologies.  The shape of $<V/V_{max}>$ is anchored by the value of $<V/V_{max}>$ at the detection limit. By contrast, the shape of theoretical log $N -$ log $P$ distributions is not explicitly anchored by the value of P at the detection limit, since different combinations of density and peak-flux could give the same limiting value of $P$ for a completely different shape of log $N -$ log $P$. Therefore, previous log $N -$ log $P$ papers have had to add extra information to constrain the shape. For example, Horack, Emslie \& Hartmann (1995) constrained log $N -$ log $P$ by using time-dilation data, Marani (1998) and Mao \& Paczynski (1992) used $<V/V_{max}>$, while Krumholz, Thorsett, \& Harrison (1998) used a total intrinsic energy.

Another novel attribute of using $<V/V_{max}>$ to trace density evolution is that plots of $<V/V_{max}>$ show relative densities directly.  Differences in densities at each redshift will cause theoretic $<V/V_{max}>$ values to diverge from observational values.  Relative density differences are harder to discern from inspection of log $N -$ log $P$ curves.

To avoid the uncertainties in the SFR, to simplify the theory, and to address more general evolutionary scenarios like SNe and quasar evolution, we adopt broken power-law models for the source density evolution function.  This approach allows us a uniform framework for addressing several different types of source evolution, without losing the character of information inherent in estimated source evolution histories.

\section{Method and Results}
\subsection{Theory}

For simplicity of calculation, we assume that GRB are standard candles with power law photon spectra $dL(E)/dE \propto E^{- \beta}$. For a source located at redshift z, the part of the spetrum  shifted into the detector energy bandwidth  ($E_{1} \leq E \leq E_{2}$) is:
\begin{equation}
L(z)= {\int_{E_1(1+z)}^{E_2 (1+z)} {dL(E)\over dE}dE}
\end{equation}
where $E_{1}$ and $E_{2}$ delimit the band-pass of detection, and L(z) is the intrinsic photon luminosity (photons sec$^{-1}$). Therefore  the observed photon flux of a GRB at redshift $z$ is (Weinberg 1972, p. 421):
 \begin{equation}
 F(z) ={L(z) /4\pi r^2(z)(1+z)} ={ \int_{E_1 (1+z)}^{E_2 (1+z)} {{dL(E)\over dE}dE} \over
          4 \pi r^2(z) (1+z) }
 \end{equation}

The additional redshift factor (1+z) is for the correction of time dilation caused by cosmology expansion. $r(z)$ is the proper motion distance, in a cosmology without cosmological constant, $r(z)$ is 
 
 \begin{equation} 
r(z)={ {c  \over H_{0}q_{0}^{2}} \left[ z q_{0}+(q_{0}-1)(-1+\sqrt {2q_{0}z+1}) \over 1+z\right ]}
   \end{equation}
$H_{0}$ is the Hubble constant, $q_{0}$ is the deceleration parameter.  The often-used statistical quantity $<V/V_{max}>$ can then be written as 
\begin{equation}
 \left< { V \over V_{max} } \right> 
             = \left< \left( { F_{min} \over F } \right)^{3/2} \right> 
    = { { \int_0^{z_{max}} \left( { F_{min} \over F } \right)^{3/2} 
                       { n(z) \over (1+z) }
                       4 \pi {r^2(z) \over \sqrt {1+H_{0}^{2} \Omega _{k} ^{2} 				  r^{2}(z)/c^{2}}}
                       { dr(z) \over dz }  
                       dz } \over
     { \int_0^{z_{max}}  { n(z) \over (1+z) }
                         4 \pi {r^2(z) \over \sqrt {1+H_{0}^{2} \Omega _{k} ^{2} r^{2}(z)/c^{2}}}
                         { dr(z) \over dz }  
                         dz } } 
 \end{equation}
where  $\Omega_{k}= -k/R_{0}^{2}H_{0}^{2}$, $k=-1, 0, 1$ for a universe that is respectively open, flat and closed, and $q_{0}=0.5 (1-\Omega_{k})$; $n(z)$ is the number of bursts per unit comoving time per unit comoving volume, and $z_{max}$ is the maximum redshift at which the dimmest bursts with $F=F_{min}$ is detected.  Therefore
\begin{equation}
N(>I)={\int_{0}^{z(I)} {4 \pi n(z)\over 1+z}{r^2(z) dr(z) \over \sqrt {1+H_{0}^{2} \Omega_{k}^{2} r^{2}(z)/c^{2}} }}
\end {equation}
\begin {equation}
I={F \over F_{min}}={{\left(1+z_{max} \over 1+z \right)}^\beta {\left(r(z_{max}) \over r(z) \right)}^2} .
\end {equation}

In a Euclidean universe, standard candle sources with a uniform spatial distribution will have $<V/V_{max}>=0.5$.  Source evolution can lead to  $<V/V_{max}>$ deviating from 0.5 (Schmidt et al.1988). In a cosmological model with constant comoving density, the effect of the universe expanding is to decrease $<V/V_{max}>$ smoothly as maximum redshift $z_{max}$ increases.  We refer this shape as a standard shape (solid line, figure 3).  For a source-evolutionary scenario, $<V/V_{max}>$ will deviate from the standard shape. Therefore, $<V/V_{max}>$ is an indicator of differences from a constant comoving source density. 

\subsection{Evolutionary Scenarios}
The comoving luminosity density of the universe, $L_{2800}$, which is considered to be a star formation indicator, has been estimated from a sample of faint galaxies over the redshift $ 0 < z < 1$ in the Canada-France Redshift Survey (Lilly et al 1996). In $q_{0} = 0.5$, $\Omega = 1 $ universe, $L_{2800}
\sim (1+z)^{3.9 \pm 0.75}$ to $z \sim 1$ and reaches its peak between 1-2.  SFR then decrease with redshift up to $z \sim 5$ (Madau et al 1996).  

A study of giant HII regions in late-type galaxies indicates that the massive stellar progenitors of Type Ib/c supernovae are probably in relatively close binary systems (Van Dyk 1996).   Type Ib/c and hypernovae progenitors have much larger masses than Type Ia progenitors, and therefore a much shorter lifetime. Their counterparts should therefore directly probe the instantaneous star formation rate.  We will therefore extrapolate the Type Ic supernovae and hypernovae formation rates from the Type II event rate and other assumptions (Madau 1998, Sadat et al. 1998).  

Given that GRB source density is proportional to its progenitor formation rate, we adopt a broken power-law evolution function as an approximation.

$n(z)={n_{0} \left( 1+z\over 1+z_{0}\right)^\alpha } \cases{\alpha =\alpha_{1} >0, z<z_{0}\cr  
                                                            \alpha =\alpha_{2} <0, z>z_{0}\cr}$

\noindent
where $n_{0}$ is the density at $z=z_{0}$, $z_{0}=1$ for SFR, Type Ic supernova and hypernova formation rate and  $z_{0}=2$ for burst sources that follow the formation rate of quasars.  

In \S 2.4 we will see that the difference between the anticipated and observed $<V/V_{max}>$ becomes larger when $\alpha_{1}$ and $|\alpha_{2}|$ increase. In order to avoid overestimating the evolution of GRBs, we choose $\alpha_{1} = 3$, medium of 3.9 for the SFR (Lilly 1996) and 2 - 2.5 for the neutron star merger rate (Totani 1997).  This parameterization can be taken as a rough description of star formation, Type Ic SN and hypernova rates.  $\alpha_{2}$ is set to be -1, 0 from the theoretical calculation for Type II SN (Madau 1998, Sadat 1998),  though the real GRB evolution might be between or stronger than these values. For burst sources that follow the formation rate of quasars, we choose $\alpha_{1} = 7$ and $\alpha_{2}= -4 $ (Yi 1994).

\subsection{ Evolution of Observed $<V/V_{max}>$ }

We selected 739 bursts from the BATSE 4B Catalog that triggered on energy channels 2+3.  Sub-samples were constructed artificially by increasing the trigger threshold  $C_{min}$ in number of counts in 64 ms and 1024ms timescale.  As $C_{min}$ was decreased incrementally by a given factor, bursts with a newly computed $C_{max}/C_{min} > 1$ were selected, $C_{max}$ is the maximum counts in 64 and 1024 ms timescale, and we choose the larger vaule of $C_{max}/C_{min}$ calculated in the two timescales. These sub-samples inherited many attributes of BATSE data although with decreasing detection rates, and hence the volumes that encompass these GRBs were incrementally increasing.  $<V/V_{max}>$ for each sub-sample is calculated by $<V/V_{max}>= <(C_{max}/k C_{min})^{-3/2}>$, where $k > 1$.

\subsection{ Results }

We set the GRB spectrum index $\beta = 1.1$ (Mallozzi et al. 1996).  The $<V/V_{max}>$ evolution with redshift in a universe with $\Omega_{\Lambda}=0$ for different evolutionary scenarios are shown in Figures 1 and 3.  $<V/V_{max}>$ for quasar-like object evolution (dash-dot-dot-dot line, $\alpha_{1}= 7$, $\alpha_{2}= -4$) shows the greatest deviation from the standard shape (solid line, $\alpha_{1} = \alpha_{2} = 0$, corresponding to non-evolving GRBs). Its peak at $z=2$ surpasses 0.6 and decreases to 0 at about $z \sim 10$, while the standard shape smoothly declines to near 0 at $z \sim 1000$. The dashed line ($\alpha_{1} =3 $, $\alpha_{2} = -1$) and dash-doted line ($\alpha_{1} =3$, $\alpha_{2} =0$) in Figure 1 are nearly unchanged at  $z<1$.  The former line slope after $z>1$ is steeper than standard shape and the later gradually overlaps the standard shape. These results show that stronger source evolution of GRBs will enlarge the deviation of $< V / V_{max}>$ from the standard shape.
The observed $<V/V_{max}> = 0.326$ for the entire 739 bursts in our sample.  This occurs at $z_{max} \sim 1.76$ for GRBs following the comoving standard shape and $z_{max} \sim 3.6$ for bursts following quasar-like object evolution.

In Figure 2, we normalize the theoretical burst number for $z_{max}$ to the number of bursts in our sample: 739.  The x-axis is the number of bursts detected for different sensitivities, selected by the method described in \S 2.3.  The y-axis is the corresponding $<V/V_{max}>$.  The dotted line is the observed evolution of $<V/V_{max}>$.  The dash-dotted line represents the $3 \sigma$ error bars obtained by the bootstrap method (Efron 1982).  The large fluctuation at the left is due to the small number in the sample.  The solid lines are the predicted distributions for different evolutionary scenarios in a flat universe.  We can see in Figure 2a that the calculated distribution with comoving constant rate can fit the observed data (dotted line) very well, while the evolution of $<V/V_{max}>$ tracing Quasar-like object rate (figure 2b) goes well beyond the $3 \sigma$ upper limit.  The intermediate parameters shown in 2c and 2d are marginally consistent with the $3 \sigma$ error line.  

In Figure 3, we compare the relations of $<V/V_{max}>$ and redshift in a flat ($\Omega_{k}=0$), open ($\Omega_{k}=0.8$) and closed ($\Omega_{k}=0$) universe for comoving constant rate (dashed lines) and quasar-like object rate (dash-dotted lines). For the observational value of $<V/V_{max}>$ (dotted line), they show a slight difference.

\section { Discussion and Conclusions }

The results presented here imply that GRB source evolution is weaker than expected for several evolutionary scenarios.  Scenarios that imply significantly fewer bright GRBs than observed, and significantly more dim GRBs, include density evolutions following SFRs, supernovae, hypernovae, and quasars. 

It is difficult to reconcile this result with indications that GRBs do occur in star-forming regions.  Several possibilities exist: the first is that GRBs, in general, do not occur at redshifts significantly in excess of $z_{0}$, where the event rate reach its peak, and indications that they do are misinterpreted.  One example of this may be GRB 971214, which has a reported redshift of $z \sim 3.4$ (Kulkarni et al. 1998).  To dramatize this, we present Figure 3, a plot of the theoretical distribution of $<V/V_{max}>$ for GRBs originating from redshifts greater than 1. The evolutionary $<V/V_{max}>$ values are clearly much higher than the observed values.  

A second possibility is that GRBs have a very wide intrinsic luminosity function (Horack et al. 1995, Krumholz, Thorsett, Harrison 1998), allowing a fraction of them to lie at very high redshifts.  If this is true, however, the dispersion of the luminosity function must not only be greater than the difference between the comoving and evolving rates, but a systematic bias must exist to shift the mean. 

A third possibility is that GRBs display a correlated density and luminosity
evolution.  One natural scenario that might allow for both types of evolution is one that admits relativistic bulk motion with a high Lorentz factor (i.e. Meszaros \& Rees 1997).  Resultant beaming creates an evolution of $<V/V_{max}>$ with redshift and increases the maximum redshift $z_{max}$ to a larger value (Che et al. 1996, Yi 1994).  This effect might shield a discrepancy caused by source evolution and perhaps adjust the $<V/V_{max}>$ distribution toward measured values (Che et al. 1998).

Lastly, we cannot ignore potential systematic biases in extrapolated star formation rates (Calzetti 1998, Cimatti et al. 1998), driven by incomplete knowledge of dust extinction, reddening, and an inability to detect low-luminosity galaxies.  Alternatively, GRBs might originate from other kinds of sources whose density and/or luminosity evolution is very different from that of known sources.

After much of this work was completed, the cited manuscripts by Lloyd \& Petrosian (1998) and Krumholz, Thorsett, \& Harrison (1998) appeared on the LANL abstract server.

This research was supported by grants from NASA and the NSF.

\clearpage
\begin{figure}
\plotone{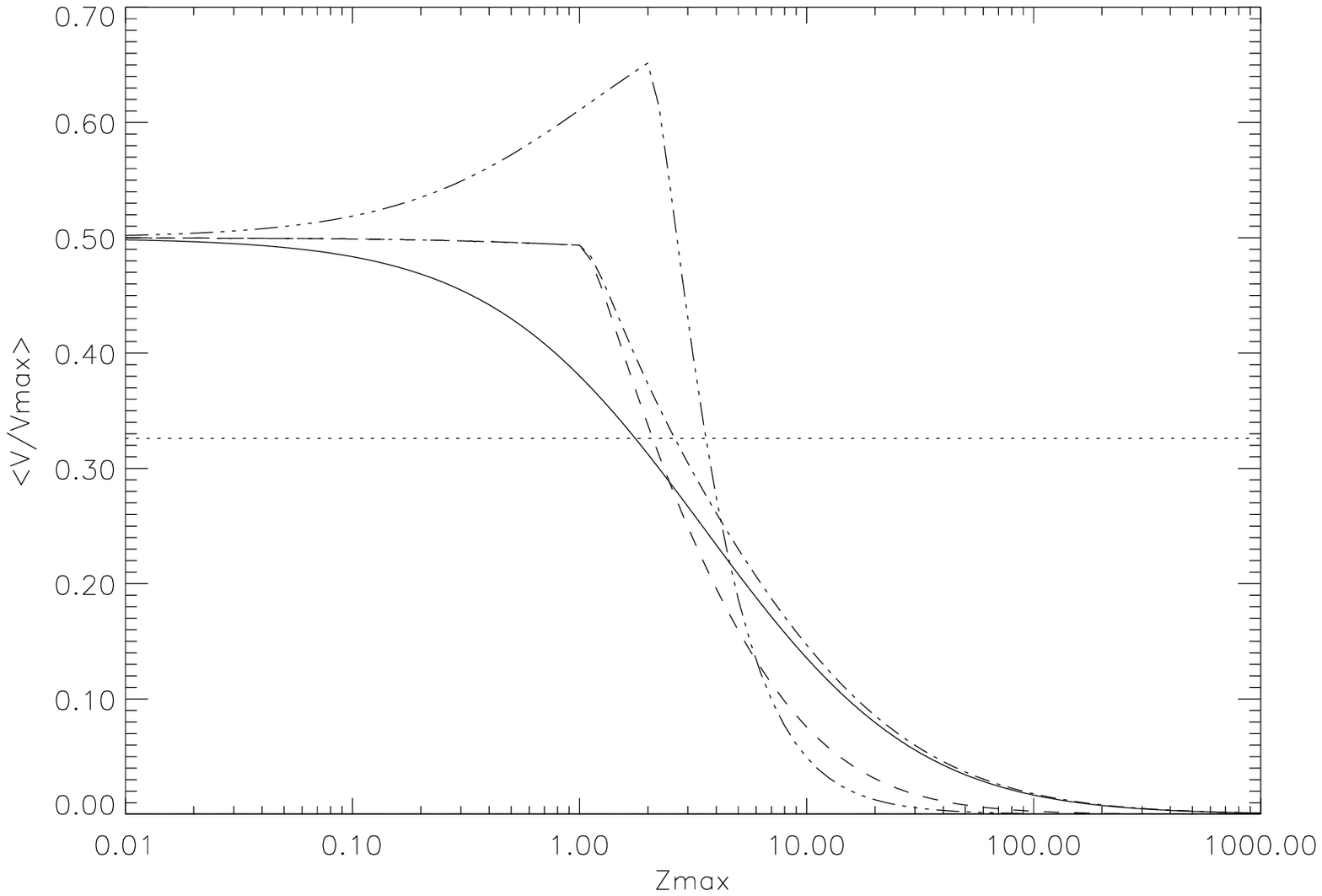}
\caption[vmax-zmax.ps]{The variation of $<V/V_{max}>$ with the maximum detection redshift $z_{max}$ is shown for the $\Omega_k=0$  universe.   GRB sources are assumed standard candles with spectral index 1.1.  The four lines (solid line, dashed line, dash-dotted line and dash-dot-dot-dot line) correspond to four evolutionary scenarios: comoving constant rate ($ \alpha_{1} = \alpha_{2} = 0 $); Type Ic supernova and hypernova scenario ($\alpha_{1}=3$, $\alpha_{2} = -1, 0$), and quasar-like object scenario
($ \alpha_{1} =7$, $\alpha_{2} = -4$). The dotted line correspond to the observed value $ <V/V_{max} >= 0.326$.
\label{fig1}}
\end{figure}

\begin{figure}
\plotone{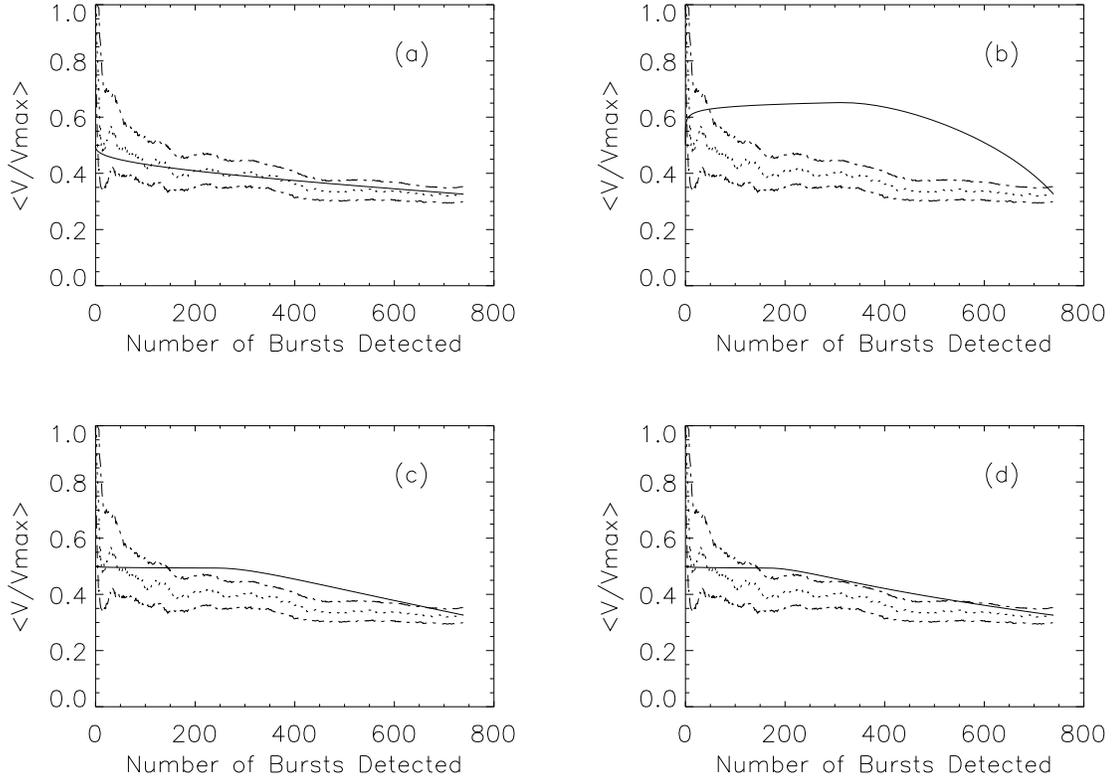}
\figcaption[compare.ps]{Comparisons of theoretical distributions of $<V/V_{max}>$ (solid lines) and the observed distribution (dotted line) obtained  from 4B catalog.  Four evolutionary scenarios are depicted: comoving constant rate (a); quasar-like object scenario (b); Type Ic supernova and hypernova scenarios (c and d) . The dotted line depicts the observed evolution. The dash-dotted lines are for the $3 \sigma $ error bars.
\label{fig2}}
\end{figure}

\begin{figure}
\plotone{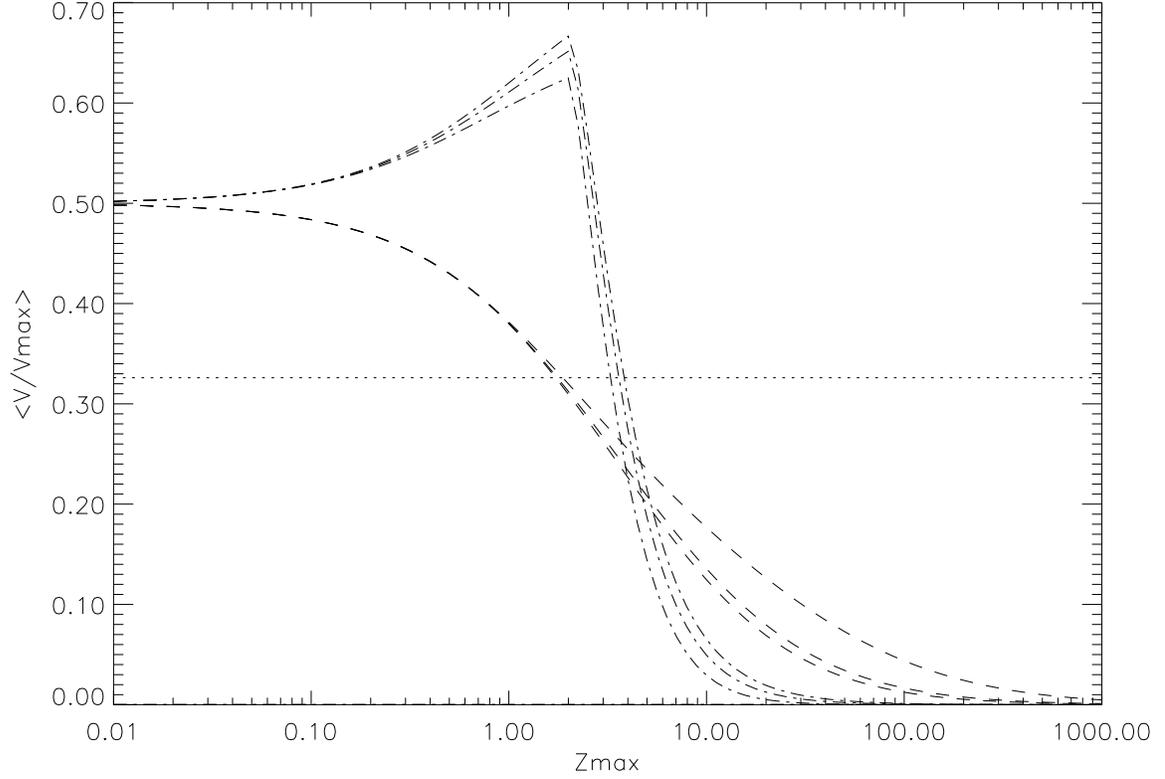}
\figcaption[cosmo1.ps]{Comparisons of evolution of $V/V_{max}$ with the maximum detection redshift $z_{max}$ in a closed ($\Omega_k=-0.8$, left line), flat ($\Omega_k=0$, middle line)  and open ($\Omega_k=0.8$, right line) universe for comoving constant rate ( dashed lines ) and quasar-like object rate (dash-dotted lines ).  The dotted line correspond to  the observed value $ <V/V_{max} >= 0.326$.
\label{fig3}}
\end{figure}

\begin{figure}
\end{figure}

\begin{figure}
\plotone{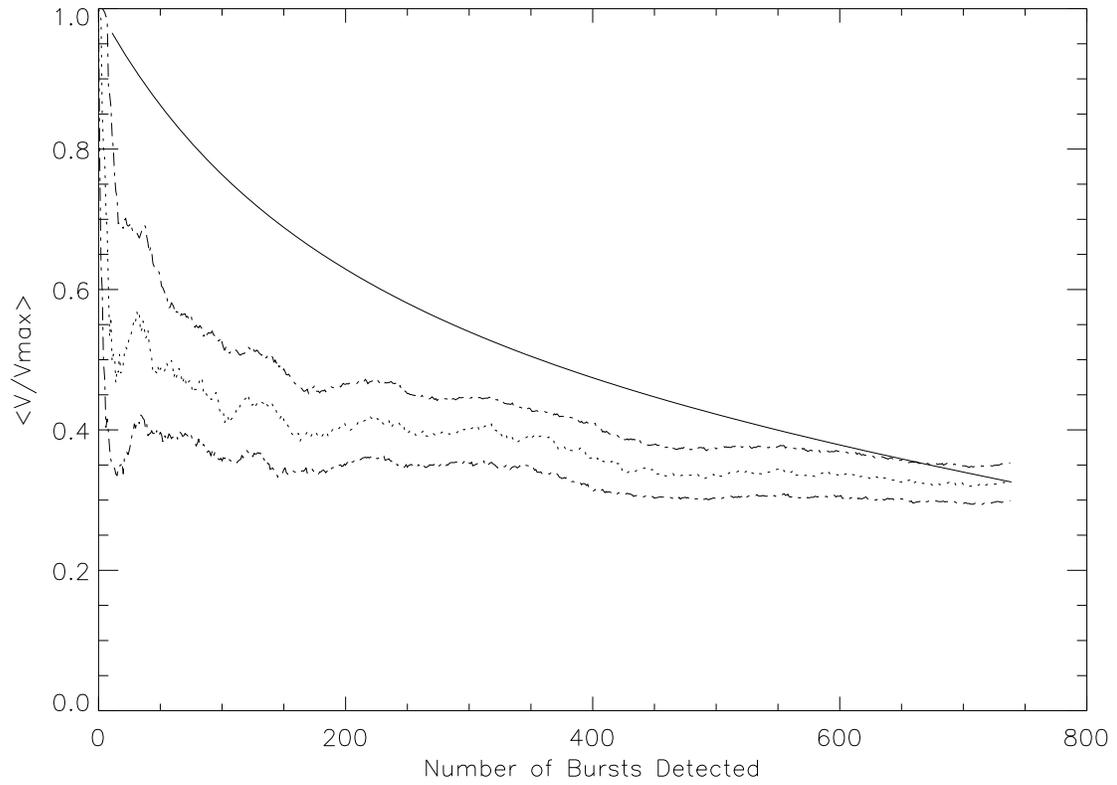}
\figcaption[vmaxz0.ps]{A comparison of theoretical distributions for GRBs originating from a redshift greater than 1. We set $\alpha_{2} =0$. The distribution for $\alpha_{2} = -1$ is similar.
\label{fig4}}
\end{figure}
\end{document}